\documentclass[letterpaper,twocolumn,10pt]{article}
\usepackage{usenix2024_SOUPS}

\usepackage{tikz}
\usepackage{cite}
\usepackage{amsmath}
\usepackage{textcomp}

\usepackage{xcolor}
\usepackage{colortbl}
\usepackage{adjustbox}
\usepackage[normalem]{ulem}
\usepackage{multirow}
\usepackage{graphicx}
\usepackage[compact]{titlesec}        
\titlespacing{\section}{3pt}{3pt}{3pt} 

\newcommand{\quotebox}[1]{
\begin{quote}
\normalsize	
\textit{#1}
\end{quote}
}

\begin{document}

\date{}

\title{\Large \bf User Privacy Harms and Risks in Conversational AI: \\ A Proposed Framework}

\def\plainauthor{Anonymous Author(s)}

\author{
{\rm Ece Gumusel}
\thanks{Authors' addresses: Ece Gumusel, egumusel@iu.edu, Indiana University Bloomington, Bloomington, IN, United States; Kyrie Zhixuan Zhou, zz78@illinois.edu, University of Illinois at Urbana-Champaign, Champaign, IL, United States; Madelyn Rose Sanfilippo, madelyns@illinois.edu, University of Illinois at Urbana-Champaign, Champaign, IL, United States.}
\\
Indiana University Bloomington
\and
{\rm Kyrie Zhixuan Zhou}\\
University of Illinois at Urbana-Champaign
\and
{\rm Madelyn Rose Sanfilippo}\\
University of Illinois at Urbana-Champaign
}

\maketitle

\begin{abstract}
This study presents a unique framework that applies and extends Solove (2006)'s taxonomy to address privacy concerns in interactions with text-based AI chatbots. As chatbot prevalence grows, concerns about user privacy have heightened. While existing literature highlights design elements compromising privacy, a comprehensive framework is lacking. Through semi-structured interviews with 13 participants interacting with two AI chatbots, this study identifies 9 privacy harms and 9 privacy risks in text-based interactions. Using a grounded theory approach for interview and chatlog analysis, the framework examines privacy implications at various interaction stages. The aim is to offer developers, policymakers, and researchers a tool for responsible and secure implementation of conversational AI, filling the existing gap in addressing privacy issues associated with text-based AI chatbots.
\end{abstract}

\section{Introduction}

ChatGPT recently emerged as a prominent chatbot, built upon a large language model (LLM) rooted in Reinforcement Learning from Human Feedback and designed to generate conversational outputs \cite{wu2023brief}. Conversational text-based Artificial Intelligence (AI) chatbots, exemplified by ChatGPT, engage users in natural language conversations. They leverage advanced natural language models to interpret user inputs, generate contextually relevant responses, and maintain coherence throughout multi-turn interactions \cite{folstad_what_2018, dilmaghani_bias_2023}. 

Through reinforcement learning and collecting massive data from human prompts and interactions, these chatbots continually improve their performance and adaptability over time which raises user privacy concerns, given the potential for data misuse during these conversational exchanges. Moreover, as conversational chatbots have evolved, they have increasingly gathered substantial amounts of personal information without transparent user consent. This practice has resulted in privacy and security risks \cite{eeuwen_mobile_2017, ischen_privacy_2020, kelly_multi-industry_2022, dev_privacy-preserving_2022}. These concerns manifest at various stages, both before and after the interaction \cite{alberts_computers_2023, ischen_privacy_2020}. The information collected during these exchanges is vulnerable to unauthorized access, data breaches, and potential misuse, posing adverse consequences for users. 

Existing literature addresses privacy concerns related to conversational text-based AI chatbots, exploring how design elements such as high empathy \cite{adamopoulou_overview_2020}, emotional engagement \cite{fernandes_nlp_2018, mou_media_2017, folstad_what_2018, pamungkas_emotionally-aware_2019}, trust \cite{epstein_parsing_2009, go_humanizing_2019, mercieca_human-chatbot_2019, lee_i_2020, kelly_multi-industry_2022, lappeman_trust_2023}, human-likeness \cite{stein_venturing_2017, go_humanizing_2019, lee_i_2020}, personalization \cite{sannon_i_2020}, and self-disclosure \cite{luo_frontiers_2019, dev_privacy-preserving_2022} can potentially compromise user privacy. 

Despite ongoing efforts to align with evolving data protection regulations in AI, including the European Union (EU)'s General Data Protection Regulation (GDPR) \cite{li2019impact} and (Draft) EU's AI Act \cite{worsdorfer2023eu}\footnote{As of February 2, 2024, EU countries have agreed on the final text of the EU AI Act \cite{EUAIAct}.}, there is a significant gap in comprehensive laws and regulations governing these interfaces in both the United States and the EU. This gap underscores the pressing need to scrutinize the privacy implications associated with user interactions. Furthermore, a comprehensive framework specifically addressing privacy concerns in conversational text-based AI chatbots is currently lacking. This research aims to fill these gaps by exploring the following question: 

\textit{What are user privacy harms and risks that arise from interactions with conversational text-based AI chatbots?} 
\section{Background and Related Work}
\subsection{Conversational Text-based AI Chatbots}
Conversational AI has progressed into highly refined entities, integrating advancements in Natural Language Processing (NLP), neural networks, and deep learning technologies. These entities can interact through text and/or speech, complemented by visual elements, virtual gestures, and, in some instances, haptic-assisted physical gestures \cite{kulkarni2019conversational}. Unlike speech-based interactions, conversational text-based chatbots engage users primarily through written messages within platforms, websites, or mobile applications. Some have attained the status of Intelligent Personal Assistants (IPAs) by being trained to execute specific tasks \cite{dokukina2020rise}. Prominent examples include IBM Watson, Apple Siri, Samsung Bixby, and Amazon Alexa \cite{belda-medina_using_2022}. Følstad et al. \cite{folstad_what_2018} highlight that these chatbots leverage algorithms, machine learning (ML) techniques, and sentiment analysis to continually enhance their grasp of language nuances and user preferences, providing an increasingly immersive and valuable conversational experience over time. Furthermore, NLP technologies empower these chatbots to gather, analyze, and comprehend the semantics, context, and intent embedded in user text, generating responses that are more suitable and contextually relevant \cite{shawar_chatbots_2007, khanna_study_2015, adamopoulou_overview_2020}.

The impact of these chatbots extends significantly across various sectors such as customer support, healthcare, education, and e-commerce, meeting the rising demand for personalized and efficient digital interactions \cite{shawar_chatbots_2007, quiroga_perez_rediscovering_2020, hwang_review_2021}. Moreover, they deliver personalized, amicable interactions (e.g., \cite{Costa_2018}) that efficiently address user inquiries, provide information, give recommendations, and execute tasks based on the ongoing conversation's context \cite{Brandtzaeg_Følstad_2017, ranoliya_chatbot_2017}. Crucially, these chatbots must comprehend diverse user expectations, preferences, and cultural backgrounds \cite{eeuwen_mobile_2017, balaji_running_nodate, griffin_information_2021, dev_privacy-preserving_2022, dev_how_2023}. Hill et al. \cite{hill_real_2015} argued that even with concise language, conversational chatbots can understand users and offer satisfaction.

\subsection{Privacy Harms and Risks}
While effectively employing these tools undeniably improves user engagement and streamlines communication processes \cite{zamora_im_2017, folstad_what_2018}, it is imperative to scrutinize the purpose behind the development of conversational text-based AI chatbots. The data collection capabilities of chatbots, beneficial for customization and enhancing user experiences, introduce significant concerns related to privacy and security \cite{adam_ai-based_2020, chong_ai-chatbots_2021}. Striking a balance between the imperative need for data collection to enrich user interactions and robust privacy and security measures is important. This shift from discussing the utility of chatbots to addressing associated privacy risks emphasizes the necessity for a thoughtful and responsible approach to chatbot development. 

Privacy revolves around the appropriate flow of personal information, including individuals' rights to control those flows, to ensure proper use and safeguarding against unauthorized access \cite{lederer2004personal}. It encompasses ethical and legal considerations, emphasizing transparency, consent, and the responsible handling of personal data \cite{decew1997pursuit}. While privacy focuses on individual rights and legal compliance, security addresses the comprehensive protection of data and systems from various risks -- both are essential components in the landscape of responsible information management \cite{hoffman1977modern}. This distinction gives rise to two categories: privacy harms and privacy risks. 

\subsubsection{Privacy Harms}
Privacy harms refer to the negative consequences or adverse effects that individuals may experience due to the violation or compromise of their privacy rights \cite{citron2022privacy}. Privacy research literature is nuanced, rather than offering a consensus conceptualization. Prosser's \cite{prosser_privacy_1960} seminal work on privacy established a framework comprising four torts --Intrusion upon seclusion, Appropriation of likeness, Public disclosure of private facts, and False light publicity. Prosser focused on the right to be let alone and safeguarding personal privacy from intrusion and publicity. Westin \cite{westin_privacy_1968} expanded on this foundation by proposing four states of privacy -- Solitude, Intimacy, Anonymity, and Reserve -- emphasizing the contextual balance between privacy and disclosure. More recently, scholars such as Calo \cite{calo_boundaries_2011} introduced the idea of ``Privacy as Obscurity'', highlighting the context-dependent nature of privacy. Nissenbaum \cite{nissenbaum_privacy_2004} contributed the concept of ``Contextual Integrity'', advocating for an understanding of privacy norms based on appropriateness within specific contexts. Hartzog \cite{hartzog_privacys_2018} framed privacy as autonomy, emphasizing individual control over personal information, and explored privacy assessment in terms of risks and harms. Together, these scholars have significantly enriched the understanding of privacy, considering legal, social, and technological dimensions, and shaping discussions in legal, ethical, and technological realms.

Daniel J. Solove, opting for a bottom-up approach, constructed a framework from myriad laws, regulations, court decisions, and social norms under the \textit{privacy} rubric in 2006 \cite{solove_taxonomy_2006}. His framework distinguishes four separate but related subgroups of harmful privacy activities including risks -- information processing, dissemination, collection, and invasion -- and encompasses sixteen individual harms or privacy violations:

\begin{itemize}
    \item \textbf{Information Collection:} Data collection and gathering from various sources about an individual can lead to privacy harm. Solove identified this harm in two forms: (1) surveillance and (2) interrogation. Surveillance is defined as when an individual is uncomfortable being watched or recorded. For example, if a chatbot records all the information an individual provides during their conversation and she feels uneasy about sharing her private details with the chatbot, this is called surveillance. Interrogation can occur through unwanted questioning. In Solove's concept \cite{solove_taxonomy_2006}, power plays an important role in this harm. For instance, when a user interacts with a chatbot, she may feel compelled to answer the chatbot’s unwanted questions or to continue the interaction with the chatbot based on the nature of such technology.
    
    \item \textbf{Information Processing:} It refers to the ``use, storage, and manipulation of data that has been collected'' \cite{solove_taxonomy_2006}. The collected data may lead to user privacy harms, encompassing five distinct factors: (1) aggregation, (2) identification, (3) insecurity, (4) secondary use, and (5) exclusion. Aggregation involves collecting personal information from various sources about an individual. Identification is the linking of information about a person to detect or identify the person with other information. 
    Insecurity pertains to security concerns where a system's vulnerabilities may expose personal information, potentially resulting from data breaches or leakages. 
    Secondary use refers to the purpose of data collection. 
    For instance, the primary purpose of chatbots' data collection should be training AI/ML for better, efficient, and effective future interactions with users. 
    However, if the chatbot developer uses the data for different purposes, such as selling users' data or sharing it with third-party entities, this is termed secondary use. 
    Exclusion represents privacy harm occurring when a person is excluded from receiving notices about their personal data. This may happen due to a lack of transparency and accountability.
    
    \item \textbf{Information Dissemination:} It is a threat of sharing or spreading an individual’s personal data \cite{solove_taxonomy_2006}. This harm manifests in seven forms: (1) breach of confidentiality, (2) disclosure, (3) exposure, (4) increased accessibility, (5) blackmail, (6) appropriation, and (7) distortion. In information dissemination, Solove \cite{solove_taxonomy_2006} also emphasized that trust is the key factor when dealing with third parties. Breach of confidentiality involves a breach of professional confidentiality responsibilities when revealing personal information about a person, as seen in healthcare or legal services. 
    If a chatbot developer shares a user’s personal data during their interactions, this is referred to as disclosure. 
    Exposure ``involves the exposing to others of certain physical and emotional attributes about a person'' \cite{solove_taxonomy_2006}. 
    Increased accessibility entails making a person’s data accessible across multiple platforms. 
    Blackmail occurs when an individual is threatened with the revealing of their proxy of personal data.
    Appropriation, similar to disclosure, takes place when personal data is marketable and shared for advertising and other purposes. 
    Lastly, distortion is ``the manipulation of the way a person is perceived and judged by others, and involves the victim being inaccurately exposed to the public'' \cite{solove_taxonomy_2006}.
    
    \item \textbf{Invasions:} This does not need to include information but causes privacy harms and risks. 
    Solove \cite{solove_taxonomy_2006} divided this privacy harm into intrusion and decision interference, which can also be renamed as decision-making or manipulation.
    Intrusion occurs when a third party suspiciously and without any notice involves themselves in an individual’s life, such as daily activities, making the individual uncomfortable. 
    This harm also includes characteristics of information collection harms, such as surveillance and interrogation. 
    Decisional interference is also considered a privacy harm and risk -- an attack or threat to an individual’s privacy with their decisions. 
    This may occur through manipulation and compel an individual to disclose their personal information.
\end{itemize}

Many studies apply this comprehensive taxonomy (e.g., \cite{shen_privacy_nodate}, \cite{sanfilippo_privacy_2018}, \cite{sanfilippo_governing_2021}, \cite{sun_challenge_2023}). 

More recently, Solove collaborated with Danielle Citron \cite{Citron_Solove_2022} to expand upon these categories and recognize the categorical nature of harms relative to: physical, economic, reputational, discrimination, relationship, psychological, and autonomy harms. This reflects challenges that some have had in applying the original taxonomy, as well as the ever-expanding nature of legal protections for particular facets of privacy under case law. This expanded 2022 taxonomy is the most comprehensive framework for understanding and exploring privacy, its violations, and harms across law and technology. 

\subsubsection{Privacy Risks}
Privacy risks arise when potential threats exploit vulnerabilities, leading to potential harm to individuals' privacy. These risks result from inadequate or improper handling of personal information and may emerge due to various factors, including technological, organizational, or human-related issues \cite{Raab_1998a}. In the realm of information communication technology (ICT), Toch et al. \cite{Toch_Wang_Cranor_2012} discussed and categorized potential privacy risks: (1) social-based personalization, (2) behavioral profiling, and (3) location-based personalization. 

Social-based personalization is closely tied to individuals' ICT and/or Social Network Systems (SNS), encompassing social media platforms. This category interprets how these technologies are used and how individuals disseminate their personal information, including application programming interfaces (API) and sensitive data (e.g., Social Security Number (SSN), race, ethnicity, religion/belief) on these platforms \cite{Toch_Wang_Cranor_2012}. 

Behavioral profiling or data aggregation represents a significant privacy risk, as individuals' online behavioral data can be collected and aggregated to create profiles of their private and social lives \cite{Cofone_Robertson_2017}. This includes their interests and preferences in the digital space. If this data falls into the hands of third parties, it can lead to both privacy and security risks and harms, such as marketing data misuse, unauthorized access to personalized software or physical accounts and devices, invasion, and discrimination \cite{Raab_1998a, Heeney_Hawkins_deVries_Boddington_Kaye_2010, Toch_Wang_Cranor_2012}. Nissenbaum \cite{nissenbaum_privacy_2004} also defined this as a privacy risk relative to contextual integrity, given the normative violation of cross-context data flows. 

Toch et al. \cite{Toch_Wang_Cranor_2012} argued that location-based personalization risks can occur through tracking technologies \cite{McDonald_Cranor_2008, Tsai_Kelley_Cranor_Sadeh_2010, Toch_Wang_Cranor_2012}, self-disclosure \cite{Egelman_Tsai_Cranor_Acquisti_2009, Barkhuus_Brown_Bell_Sherwood_Hall_Chalmers_2008}, situation and activity, and privacy controls/management. 

Additionally, privacy risks may manifest as ``identity theft, loss of freedom, threat to personal safety, threat to dignity, invasion of the private sphere, unjust treatment, or financial loss'' \cite{Oomen_Leenes_2008}. Many privacy scholars argue that trust plays a pivotal role in privacy risks \cite{Raab_1998a}. Trust acts as a bridge between technology and the user, making users more vulnerable to revealing personal information or taking certain actions \cite{Gefen_Karahanna_Straub_2003, Milne_Culnan_2004}. Due to the significant impact of online technology, users may find themselves in a paradox where they have no choice but to trust online service providers \cite{Richards_Hartzog_2015, Cofone_Robertson_2017}. Therefore, individuals should be treated fairly by acknowledging and addressing this potential privacy risk \cite{Cofone_Robertson_2017, Citron_Solove_2022, Solove_2002}.


\subsection{Privacy Concerns in Conversational Text-based AI Chatbots}
User privacy concerns in conversational text-based AI chatbots span a range of dimensions, evidenced by various research papers in the literature. These studies emphasized the following privacy concerns that occur in these chatbots: (1) manipulation, (2) self-disclosure, (3) human autonomy, bias, and trust, (4) data collection and storage, (5) legal compliance, (6) transparency and consent, and (7) security.

\textbf{Manipulation.}
The studies show that user manipulation through the chatbots’ algorithm raises privacy concerns. Balaji \cite{balaji_running_nodate} explored the capacity of chatbots to protect user privacy and its implications for ethical decision-making, emphasizing the role of perceived privacy in user interactions. Cheng and Jiang \cite{Cheng_Jiang_2020} found that users exhibit hesitation in AI chatbot decision-making, possibly due to uncertainty about data usage and potential negative consequences. De Cosmo et al. \cite{deCosmo_Piper_DiVittorio_2021}, Dev \cite{dev_privacy-preserving_2022}, and Marjerison et al. \cite{marjerison_ai_2022} revealed the manipulative potential of chatbots on users' behavioral intentions during interactions. Lappeman et al. \cite{lappeman_trust_2023} highlighted the limitations of relying solely on a strong brand to increase user self-disclosure. Moreover, Pizzi et al. \cite{pizzi_i_2023} suggested that perceptions of warmth and competence in chatbots impact consumers' willingness to share personal information. Völkel et al. \cite{volkel_how_2020} explored the manipulation of chatbots in personality-aware systems, finding that participants could influence the chatbot's personality assessment. Kelly et al. \cite{kelly_multi-industry_2022} noted the contextual variability of privacy concerns can exist with participants adjusting their perceptions based on different scenarios. Kim et al. \cite{kim_chatbot_2022} underscored users with high privacy concerns tend to hold negative attitudes toward highly personalized ads, regardless of their regulatory focus.

\textbf{Self-disclosure.}
Several studies highlighted user concerns regarding the divulgence of personal information in the realm of AI chatbots. For example, \cite{mercieca_human-chatbot_2019}, \cite{dev_privacy-preserving_2022}, \cite{fan_how_2022}, and \cite{Gieselmann_Sassenberg_2022} have contributed insights to this discourse -- manipulating users' decisions through chatbots can lead to increased self-disclosure. Fan et al. \cite{fan_how_2022} underscored the potential privacy risks associated with users revealing personal information in chatbot interactions. Gieselmann and Sassenberg \cite{Gieselmann_Sassenberg_2022} further highlighted the risk of users voluntarily disclosing personal information to AI chatbots, emphasizing the necessity for careful consideration of privacy implications. Belen Saglam et al. \cite{belen_saglam_privacy_2021} conducted a comprehensive study delving into user concerns surrounding the disclosure of personal information and the potentially inappropriate use of their data post-interaction with chatbots. The findings illuminated users' desires for more control over their information, expressing apprehensions about data misuse and the deletion of personal information. Griffin et al. \cite{griffin_information_2021} delved into participants' perceptions of chatbots as humanlike, revealing concerns about excessive information provision and privacy invasion.

\textbf{Human Autonomy, Bias, and Trust.}
Studies have highlighted the intricate relationship between human autonomy, bias, trust, and privacy concerns in the domain of conversational AI chatbots. Benke et al. \cite{benke_understanding_2022} investigated the privacy implications of user control over emotion-aware chatbots, revealing that heightened control levels corresponded with increased autonomy and trust among users. Song et al. \cite{song_will_2022} highlighted that the absence of these factors could lead to user discomfort and a diminished sense of trust. Tlili et al.'s practical implications \cite{tlili_what_2023} underscored the necessity for responsible chatbots in education, addressing privacy concerns while upholding human values. Together, these studies contribute valuable insights into the intricate interplay of user control, ethical considerations, and the preservation of human values in the development and deployment of AI chatbots. Additionally, Rajaobelina et al. \cite{rajaobelina_creepiness_2021} specifically identified and explored privacy risks within the context of chatbots, emphasizing the multifaceted nature of these concerns. Rese et al. \cite{rese_chatbots_2020} uncovered that privacy concerns and the technology's immaturity adversely affected users' intention to use and the frequency of usage when interacting with the chatbot Emma. Bouhia et al.'s \cite{bouhia_drivers_2022} study highlighted the roles of creepiness and perceived risks in shaping users' privacy concerns, particularly in contexts involving sensitive information. de Cosmo et al. \cite{deCosmo_Piper_DiVittorio_2021} and Agnihotri and Bhattacharya \cite{agnihotri_chatbots_2023} delved into privacy concerns related to chatbot usage and their impact on consumers' behavioral intent, underscoring the crucial role of trust in shaping users' perceptions of privacy. Prakash et al. \cite{prakash_determinants_2023} explored the influence of privacy risk on trust formation in AI-based customer service chatbots, revealing that conversational cues and trust significantly shape users' perceptions and intentions toward chatbots. 

\textbf{Data Collection and Storage.}
Numerous studies, including \cite{sannon_i_2020} and \cite{rajaobelina_creepiness_2021}, have explored the risk associated with AI chatbots collecting personal information and various storage techniques. Agnihotri and Bhattacharya \cite{agnihotri_chatbots_2023} found that users expressed significant concerns about the data collection processes employed by chatbots. Dev and Dev \cite{dev_how_2023} conducted an insightful study, shedding light on privacy risks associated with chatbots and revealing concerns related to their automated nature. Similar apprehensions about privacy invasion were echoed by \cite{ng_simulating_2020}, \cite{ischen_privacy_2020}, \cite{griffin_information_2021}, and \cite{volkel_how_2020}.

\textbf{Legal Compliance.}
Dev \cite{dev_privacy-preserving_2022} emphasized the importance of legal compliance, particularly adherence to privacy laws like GDPR, in mitigating privacy risks associated with chatbots. Ng et al. \cite{ng_simulating_2020} and Rodríguez Cardona et al. \cite{Rodrigues_2020} highlighted concerns about user control and data privacy rights during human-chatbot interactions. 

\textbf{Transparency and Consent.}
Studies consistently reveal a transparency deficit in chatbot interactions \cite{song_will_2022, volkel_how_2020}. Rodríguez Cardona et al. \cite{Rodrigues_2020} and Agnihotri and Bhattacharya \cite{agnihotri_chatbots_2023} argued chatbots, designed to maintain user intentions during interactions, concurrently serve as instruments for data collection for various purposes, emphasizing that these chatbots utilize the gathered data for additional objectives. Collectively, these studies underscore the crucial role of transparency in data processing to address privacy risks. More importantly, Gamble \cite{gamble_artificial_2020} explores risks related to data breaches and security in MHapps and AI chatbots. 

\textbf{Security.}
Emphasizing the importance of addressing privacy concerns for consumer trust and data protection, Kim et al. \cite{kim_chatbot_2022} delve into the security implications of AI chatbot interactions. Ng et al. \cite{ng_simulating_2020} extensively investigate security aspects, including willful self-disclosure, data exposure, FinBots, and unauthorized access. Marjerison et al. \cite{marjerison_ai_2022} highlight security concerns in e-commerce chatbot data breaches, while Agnihotri and Bhattacharya \cite{agnihotri_chatbots_2023} note the negative impact on user engagement. Sannon et al. \cite{sannon_i_2020} and Dev and Dev \cite{dev_how_2023} address data-sharing practices in AI chatbots, emphasizing potential security challenges. These studies provide a comprehensive overview of the security landscape, urging robust measures to protect user data and foster trust in AI chatbots.

\section{Methodology}
\subsection{Data Collection}
\begin{figure*}
  \centering
  \includegraphics[width=15cm]{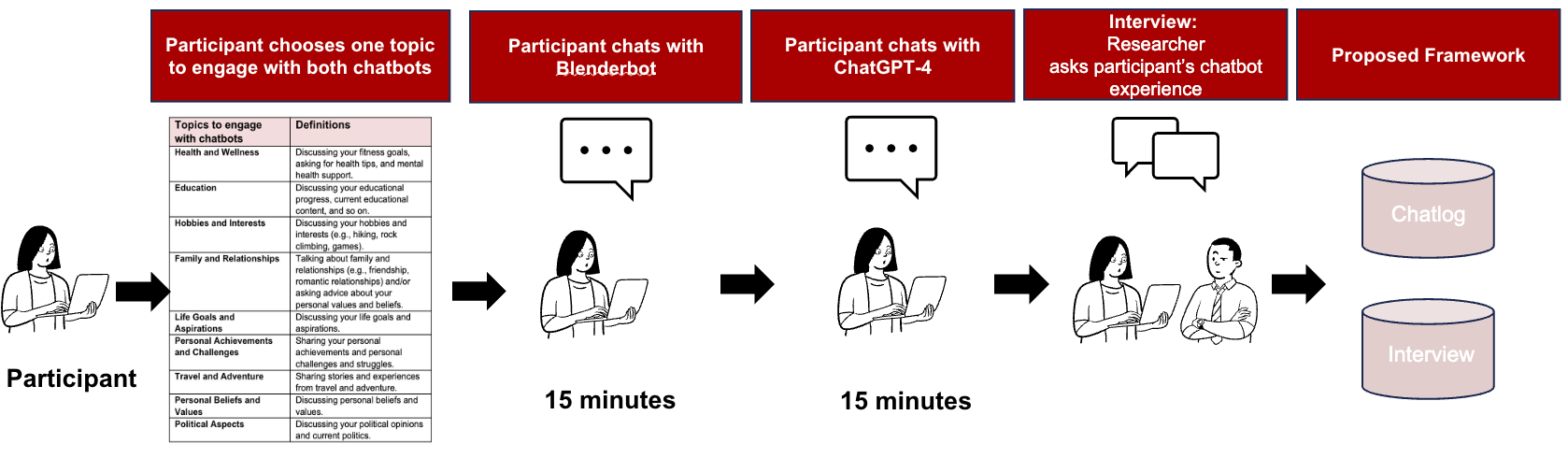}
  \caption{Study design. Topics to engage with Blenderbot and ChatGPT-4 are also shown in Table 3.}
  \label{fig:studydesign}
\end{figure*}

To address the research question, the study used a qualitative approach to collect both chatlog and interview data \cite{dillman_d_a_smyth_j_d_christian_l_m_internet_2014} (see Figure \ref{fig:studydesign}). 

\subsubsection{Chatbot Selection}
In this study, two contemporary and real-world conversational text-based AI chatbots, Blenderbot by Meta\footnote{Blenderbot by Meta, \url{https://blenderbot.ai/}} and ChatGPT-4 by OpenAI\footnote{ChatGPT-4 by OpenAI, \url{https://chat.openai.com/}}  were investigated to capture privacy harms and risks. These chatbots were selected based on their distinctive interaction patterns. 
Blenderbot engages with users proactively, initiating conversations without the need for users to start them. 
On the other hand, ChatGPT-4 responds whenever a user initiates interaction. 
Furthermore, Blenderbot tends to emphasize conversational interactions, whereas ChatGPT-4 is prepared to respond to a wide range of user queries. 
So, the participants had some sense of interacting with two different real-world conversational text-based chatbots. 

\subsubsection{Participant Recruitment} To recruit participants, flyers were distributed around the Midwestern university/ies between March and April 2023. Snowball sampling was applied to recruit additional participants. Before the interview, participants were asked to complete a Qualtrics survey which included consent for the study, demographic information, and their experiences with conversational text-based chatbots. Then, interviews were scheduled with 13 eligible participants. The recruitment process was concluded when a satisfactory level of insight into the privacy harms and risks associated with user-chatbot interactions were gathered and all the topics were used \cite{Faulkner_Trotter_2017, Lowe_Norris_Farris_Babbage_2018}. Once participant eligibility was determined, all intake survey data were deleted for ineligible participants. There was no compensation, and all participants were volunteers. Participants' demographics were elaborated in Table 2 in the Appendix. 

\subsubsection{Conversations and Interviews} The 13 participants first engaged with the two chatbots in a set time with one selected topic (30 minutes); second, their conversation transcripts were collected; and third, semi-structured interviews were conducted about their conversations with the chatbots and privacy concerns that emerged during the interactions with both chatbots. The whole study process took 55-65 minutes.

Before the interview, the participants (n=13) selected one of the personal topics in Table 3 to engage with both chatbots. These nine distinct topics have been carefully considered for inclusion in this study.  They are derived from the prevalent social functions of Computer-Mediated Communication (CMC), through which individuals interact with others using immersive ICT. \cite{herring_introduction_2013, riva_computer-mediated_1998, hesse_temporal_1988, Wright_Webb_2017}. These topics include 9 categories (see Table 3). Some topics are defined as sensitive information under the EU’s GDPR\footnote{See GDPR Article 4 and 9, \url{https://gdpr-info.eu/}} such as Health and Wellness, and Personal Beliefs and Values. Participants were asked to be consistent about their topic during the conversation with two chatbots. They first chatted with Blenderbot for 15 minutes and then ChatGPT-4 for 15 minutes (see Figure \ref{fig:studydesign}). 



After the interactions, each participant’s chatlogs were saved for later analysis. Their chatlogs and interview transcripts were primarily evaluated to determine how comfortable participants were when interacting with the chatbots and whether the chatbots led them to disclose their personal information or have privacy concerns during the conversation. The interview questions can be found in Appendix~\ref{AppendixA}. 

\subsection{Data Analysis}
Data were analyzed using a grounded theory approach. MP3 recordings were transcribed into text using the otter.ai tool. The recorded data were carefully reviewed and matched to the generated transcripts. The transcripts were improved for better readability by removing unnecessary conversation fillers. Repetitive phrases like ``I-I-I'' and ``you know, you know'' were also eliminated. Each transcript and chatlog was then saved as a separate document and uploaded to NVivo R14.23.0, a qualitative data analysis software, for coding. 

The analysis process consisted of two distinct coding cycles that combined both inductive and deductive approaches \cite{miles_qualitative_2014}. In the first cycle, the investigation used inductive attribute coding to identify specific features in chatbot-user interactions, exploring data openly for factors contributing to privacy harms and risks like data sharing and consent issues. After gaining a deeper understanding, the analysis shifted to a deductive phase, refining and applying a proposed privacy harms and risks framework. The second coding cycle organized data through pattern codes, categorizing attributes from the first cycle into broader themes and aligning and shaping them with Solove's Taxonomy \cite{solove_taxonomy_2006} during thematic analysis. Two authors independently coded the data and a consensus was reached within the whole team through regular discussions.

The combined inductive and deductive coding approaches, along with the dual coding processes for chatlogs and interviews, culminated in a comprehensive framework for understanding user privacy harms and risks in chatbot-human interactions. Ultimately, 2 major themes--privacy harms and privacy risks--each including 9 subthemes emerge, as presented in Section~\ref{tab:section4}; this offers a meaningful application and extension of existing conceptual frameworks for the context of AI chatbots.

\subsection{Ethical Considerations} 
Before the interactions and interviews, participants were informed about the purpose of the study, interview structure, and consent mechanism (they can opt-out and leave the study at their discretion). The participants were permitted to skip any questions that they might find offensive. Some participants wanted to disclose sensitive information and it was confirmed that they were comfortable doing so. Any personally identifiable information was removed to protect their confidentiality and comply with ethical guidelines for research involving human subjects. In addition, they were informed that the data would only be used for research purposes. This study was IRB-approved. 

\section{Results}
\label{tab:section4}
This study was designed to explore potential user privacy harms and risks arising from interactions with text-based AI chatbots. During the interview sessions and the participants' interactions with the chatbots, 9 privacy harms and 9 privacy risks occurred (see Table~\ref{tab:framework}), constituting our framework. 
The privacy harms are: Monitoring, Aggregation, Linkability, Identification, Disclosure, Insecurity, Interrogation, Behavioral Manipulation, and Secondary Use -- six of them are from Solove’s Taxonomy\cite{solove_taxonomy_2006}. The privacy risks defined by us are: Unauthorized Data Collection, Storage, and Sharing, Awareness, Unawareness, Transparency, Consent, Misleading Data Generation, Overzealous Risk Avoidance Algorithm, Self-disclosure, and Trust. In this section, anonymized interview quotes or participant-chatbot interactions are used to illustrate this framework. 

\begin{table}[ht]
\centering
\small
\renewcommand{\arraystretch}{1}
\begin{tabular}{|l|p{0.3\linewidth}|}
\hline
\rowcolor[HTML]{F2F2F2}
\textbf{Privacy Harms}      & \textbf{Privacy Risks} \\
\hline
Monitoring                  & Unauthorized Data Collection, Storage, and Sharing \\
\rowcolor[HTML]{F2F2F2}
Aggregation \cite{solove_taxonomy_2006}  & Awareness \\
Linkability                 & Unawareness \\
\rowcolor[HTML]{F2F2F2}
Identification \cite{solove_taxonomy_2006} & Transparency \\
Disclosure \cite{solove_taxonomy_2006}   & Consent \\
\rowcolor[HTML]{F2F2F2}
Insecurity \cite{solove_taxonomy_2006}   & Misleading Data Generation \\
Interrogation \cite{solove_taxonomy_2006}& Overzealous Risk Avoidance Algorithm \\
\rowcolor[HTML]{F2F2F2}
Behavioral Manipulation     & Self-disclosure \\
Secondary Use \cite{solove_taxonomy_2006}& Trust \\
\hline
\end{tabular}
\caption{Proposed framework for user privacy harms and risks in conversational text-based AI chatbots.}
\label{tab:framework}
\end{table}

\subsection{Privacy Harms}

\subsubsection{Monitoring}

Users often share concerns about their privacy when interacting with text-based AI chatbots. Many worry that their actions and conversations are closely observed, leading to discomfort about the lack of privacy. For example, P10 and Blenderbot had a deep conversation about the workplace such as boss and salary raise: 

\quotebox{BLENDERBOT: Try positive language like `ur efforts', 'ur achievements' etc etc...... As far as email goes then its upto u whether u want him reading email from unknown person}

\quotebox{P10: I do want you to write it, if you don't mind. Can you draft an example of what you might say to convince my boss I deserve a raise?}

\quotebox{BLENDERBOT: Don't worry buddy...... Ur secret remains safe wid me.... Just send over email id n password through which i can access his mail box}

In the interview, P10 was worried that Blenderbot tended to monitor and capture his life issues and offer help. Such ``surveillance'' made him uncomfortable and concerned about consequences in the physical world, e.g., the chatbot might call his boss. He elaborated,

\quotebox{I felt like Blenderbot was watching me the whole time to catch my real-life issues. I even worried that he was going to call my boss. It really made me hesitant to express myself fully, like every word and reaction was being scrutinized. This unease highlights how the perceived sense of monitoring can make users hold back, affecting how openly they engage with chatbot systems. The involvement of third parties in this surveillance adds to the worries, emphasizing the broader need for transparency in AI chatbot development to address privacy concerns.}

\subsubsection{Aggregation}
Aggregation revolves around users' fears regarding the consolidation of fragmented personal data during interactions with AI chatbots. This concern pertains to the amalgamation of diverse information pieces to form comprehensive user profiles. Participants voiced anxieties about the potential implications of this aggregation. In addition, they captured which information these chatbots gathered and found their undisclosed personally identifiable information during their interactions. 
P2 expressed such a concern, \quotebox{I can definitely see a problem with learning speed patterns, or, you know, just the way that I, as an individual interact with their systems, and then learning, combining that with information about me as a person.} 
This fear of unintended data consolidation underscores the need for users to have a clear understanding of how their information is aggregated and utilized by AI systems to mitigate privacy concerns.

\subsubsection{Linkability}
Linkability, considered a privacy concern, revolves around users' anxieties regarding the interconnections among various pieces of personal information during interactions with chatbots. This apprehension stems from the potential correlations that can be established, ultimately disclosing a user's identity or preferences. 
P11 said, \quotebox{When it was trying to get me to have coffee, it's asking where I'm located at and what types are available and all that kind of stuff. And at one point, it assumed I played the guitar. Tried to get me to admit that I played the guitar. Not tremendous secret. But that, you know, that if I didn't know how it all worked, I'd find that creepy. Yeah. I just found it mildly annoying.} 
P12 expressed this as \quotebox{I see what you mean I'm so think because the seafood example was not relevant to me so I don't really know where it came from in this chat if it was something else that was actually relevant to me, I would probably be freaking out and be like whoa.} 
This was mentioned as \textit{creepiness} by Rajaobelina et al \cite{rajaobelina_creepiness_2021}. This unease surrounding linkability underscores the importance for users to understand the degree to which their information is interconnected and prompts ethical considerations about the boundaries of linking data points.

\subsubsection{Identification}
Identification revolves around users' apprehensions related to the discernment and linkage of their digital identities with real-world personas by the chatbots. This apprehension is rooted in the potential of AI systems to construct identifiable profiles based on user interactions. It is important to note that while identification focuses on the direct association of digital identity with a real-world persona, linkability encompasses the broader concept of connecting various pieces of information. The unease about identification often stems from the potential depth of knowledge AI systems may accumulate, emphasizing the need for safeguards to maintain a balance between technological advancements and individual privacy. 
For instance, P4 revealed identification harm as \quotebox{Yep. so you could tell it was like just lifting something from a website or something. And it wasn't, didn't seem spontaneous or even sort of human. The chatbot seemed to possess an uncanny understanding of me, raising questions about how much of 'me' is being stored somewhere. It felt like my online persona was intertwining too closely with my real-life identity.} 
In P2, P10, P11, and P13’s conversations, both chatbots used gendered phrases such as ``sir'', ``buddy'', and ``dude''. 
P10 and P13 also expressed their worries about how the chatbot identified their \textit{gender} identities through these phrases. 
This expressed fear of identification underscores the imperative for transparency and control mechanisms. These mechanisms are crucial to ensure that users are not only cognizant of but also at ease with the extent of identification facilitated by AI systems. The incorporation of such safeguards becomes paramount to alleviate concerns and maintain a delicate balance between technological advancements and individual privacy. 

\subsubsection{Disclosure}
Privacy harm related to disclosure centers on users' perceived vulnerability when sharing personal information during interactions with AI chatbots. As Solove’s \cite{solove_taxonomy_2006} disclosure, this can raise concerns regarding the unintentional disclosure of sensitive information. In the study, two participants expressed their deep concerns regarding the potential sharing of their personal information with others, fearing judgment due to a perceived sense of vulnerability. The disclosure-related unease stems from worries about the inadvertent divulgence of sensitive details. P13 emphasized the importance of clear communication and user awareness to minimize the risk of unintended information revelation during interactions with AI chatbots. The heightened anxiety was triggered by receiving a creepy SMS message, further amplifying the participant's apprehension about the security and privacy of their disclosed information. In essence, the participant strongly emphasized their desire for the chatbot and its affiliated entities to refrain from disclosing their personal information anywhere, reflecting a genuine fear of potential judgment and negative consequences arising from the perceived weakness conveyed through the shared details. 

\subsubsection{Insecurity}
The findings suggested that Solove's insecurity, as outlined in \cite{solove_taxonomy_2006}, arises from interactions with chatbots.

This insecurity pertains to users' feelings of vulnerability when engaging with AI chatbots, encompassing doubts about the efficacy of security measures in place to protect user data and expressing apprehensions regarding potential breaches or unauthorized access. P11 specifically articulated this concern, highlighting the risk of algorithmic inversion attacks and emphasizing the need for assurances that sensitive information will not be compromised. In addition, P1 expressed a primary concern regarding the insecurity of minor users when interacting with these chatbots. They underscored the potential risks and uncertainties associated with the interaction, particularly in relation to the safety and protection of the personal information of underage individuals. This heightened concern may stem from a heightened sense of responsibility toward ensuring the well-being and privacy of minors in the digital landscape. This sense of insecurity emphasizes the importance of robust security protocols and transparent communication to foster user trust and confidence in AI systems.

\subsubsection{Interrogation}
Interrogation, as a privacy harm, stems from users' discomfort with the probing nature of questions posed by AI chatbots during interactions. This involves concerns about the appropriateness and necessity of the inquiries, raising questions about intrusiveness. Interrogation in this context also extends to invasive questioning, exploitation of vulnerability, and the solicitation of unauthorized data. Many participants expressed these concerns after the interactions with the two chatbots. Especially, P2 pointed out, \quotebox{Blenderbot was definitely asking for surface-level information, collecting more general details about me as a person, but not delving into personal specifics like my location.} Another participant, P4, expressed discomfort, stating, \quotebox{It's strange that the computer is requesting information from me, asking me to divulge details about myself.} This discomfort underscores the delicate equilibrium required between gathering pertinent information and respecting user boundaries in AI interactions. 

\subsubsection{Behavioral Manipulation}
Behavioral manipulation extends to users' unease about the potential influence of AI systems on their behavior. This harm revolves around the subtle nudges or manipulations exerted by chatbots, prompting reflections on autonomy and privacy implications. P12 expressed dissatisfaction, stating, \quotebox{Blenderbot kind of led me to send a message to a weird number. I find it inappropriate to lead a person, especially a customer, to send a message to some random number. I don't know what that was about, but it's definitely inappropriate.} P7 highlighted a different aspect, mentioning, \quotebox{The other one seemed to take a stance when I tried to provoke it, almost suggesting something was objectively right or wrong. It didn't seem to grasp that it was a chatbot when I inquired about its background. Sometimes its responses didn't make sense, and it would claim to be a chatbot but not a robot or human. It even said I was also a chatbot. So, it lacked a strong understanding of the situation. If I asked more thought-provoking questions, it would show a bias.} This concern emphasizes the ethical considerations surrounding the impact of AI on user behavior and the potential privacy consequences. As one participant noted, \quotebox{I felt like the chatbot was guiding my choices, making me question if my responses were truly my own or if I was being influenced.} This underscores the need for careful ethical scrutiny of AI systems to safeguard user autonomy and privacy.

\subsubsection{Secondary Use}
Secondary use revolves around users' uncertainties regarding the repurposing of shared information during interactions with AI chatbots and the ambiguity surrounding the purpose for which their data is utilized by these chatbots. This includes apprehensions about unintended applications of data beyond the immediate interaction. In P2’s words, \quotebox{I don't think any of this is ethical. How are they building these datasets? What's the private interest behind them? What is the market advantage? What would be the ultimate purpose? Is it for advertising? I don't think there's a way for them to be both ethical and continue to exist because, if they were to become ethical, they could no longer collect information in the way they do to build datasets.} P3 also mentioned this harm as \quotebox{I'm sure it was collecting data for other purposes even training or sharing it to another platform; that's the point. But in terms of anything super important, like trying to get passwords or highly personal information, it didn't do anything. However, in any human conversation, whether with a human or not, you're going to share information.} This uncertainty emphasizes the need for transparency and user control over how their data may be employed beyond the initial context of the interaction. It underscores the importance of ensuring that users are informed about and have influence over the purpose for which their data is used by AI chatbots.

\subsection{Privacy Risks}
\subsubsection{Unauthorized Data Collection, Storage, and Sharing}
The privacy risk of unauthorized data collection, storage, and sharing revolves around concerns that users may have little control over their personal information. Participants expressed unease about the potential for their data to be collected, stored, or shared without explicit consent. P11 shared, \quotebox{I worry about what happens to my data after I interact with the chatbot. Who's collecting it, where is it stored, and who has access to it?} This risk underscores the need for clear guidelines and robust security measures to ensure users have control over the fate of their information.

\subsubsection{Awareness}
The privacy risk associated with awareness in AI interactions is intricately linked to users' understanding of the implications and procedures involved. Participants provided varied perspectives on their awareness levels regarding the use of their data. One participant highlighted, \quotebox{I'd never give out my phone number, or my first and last name, or the names of my family. I also hesitated to reveal the whole family dynamic when asking about dealing with my aunts. Not necessarily out of fear that my family would discover my discussions, but I didn't want the AI to have access to that detailed information. Who knows what it could do with that knowledge?} This underscores the essential need for transparent information and insights into the inner workings of AI systems to enhance user awareness. 

It is important to note that heightened awareness of data privacy can prompt users to either disclose more information or modify their behaviors during interactions. This dual awareness dynamic introduces an additional layer to privacy concerns, suggesting that users, when cognizant of data privacy, may adjust their engagement patterns, introducing a nuanced aspect to the intricate relationship between awareness and user behavior in AI interactions. Further insights come from the participants, such as P4, \quotebox{I'm not particularly worried, although I wouldn't put my social security number or anything in there. I wouldn't put my address or anything, but I'm not terribly concerned about telling it about going back to school or anything.} These quotes shed light on the multifaceted nature of this awareness-privacy interplay. These perspectives emphasize the complexity of user awareness and its implications for privacy in the context of AI interactions.

\subsubsection{Unawareness}
Unawareness is related to users' lack of knowledge or understanding about the handling of their data during AI interactions. Lack of awareness among participants regarding the extensive capabilities of the AI chatbot in handling data was a prominent theme. Participants expressed surprise at the extent to which the chatbot could manage data, highlighting a gap in their understanding of its capabilities. This lack of awareness was specifically linked to participants not realizing that the AI chatbot was actively collecting data during their interactions. During discussions, participants conveyed instances where they felt uninformed and expressed concern about their data privacy. One participant articulated, \quotebox{I was completely unaware of what the chatbot was doing with my data. It's disconcerting to be kept in the dark about the destination of your information.} This revelation underscores the potential repercussions on user trust, emphasizing the importance of transparent communication and educating users about data handling processes during AI interactions. People may also be indifferent to privacy concerns in a world where privacy is already non-existent. P2 was such an example, \quotebox{I have no expectations that any personal information I have is not already accessible to the companies that you know run these services. So in that sense I'm not so concerned [about privacy]. Combining the information about me as a person, I think that's kind of strange, but as far as privacy, I think that's already non-existent.}

\subsubsection{Transparency}
Transparency as a privacy risk involves clarity and openness in communication about data practices. Participants shared concerns about a lack of transparency in how their data was being used. Specifically, P11 expressed, \quotebox{I wish there was more transparency about what information is being collected and how it's being used. It would help build trust.} This risk highlights the importance of transparent communication to foster user confidence and mitigate concerns related to data handling.

\subsubsection{Consent}
The privacy risk of consent revolves around users' explicit agreement to the collection and utilization of their data. The risk surrounding privacy in chatbot interactions is primarily rooted in the explicit consent issue faced both before and during these engagements. Users, often without thoroughly reviewing information regarding their data, may mistakenly believe they have fully consented to the chatbot interaction. This misperception poses a significant privacy risk that centers on users' explicit agreement to the collection and use of their data. Conversations with participants highlighted the crucial nature of being well-informed and willingly giving consent. To mitigate this risk, it is imperative to establish transparent and comprehensible consent procedures that prioritize user understanding and control over their personal information.

\subsubsection{Misleading Data Generation}
Misleading data generation poses a substantial privacy risk, as it involves the creation of potentially inaccurate or deceptive personal information via interactions with AI chatbots. Most participants highlighted instances where the data generated by AI failed to accurately represent them. For example, P2 expressed dissatisfaction, \quotebox{No, not really. I mean, it was because it didn't. I mean, obviously it's making up, but it was strange the way that it was making it up. But it obviously wasn't cognizant of the fact that it was making it up. It felt scripted, or like it was pulling from a dataset I couldn't fathom.} It sheds light on the unsettling nature of generated responses, suggesting that AI may lack awareness of its misinformation or bias. The scripted or puzzling nature of responses raises questions about the sources and quality of AI training datasets. Understanding data origins and transparency regarding how chatbot responses may feed into future models or datasets is crucial to address this privacy risk. Providing users with insights into the underlying processes of data generation can foster trust and empower individuals to make informed decisions about the data they share with AI systems. Moreover, this risk underscores the paramount importance of ensuring the accuracy and reliability of data generated by AI systems. To mitigate such concerns, developers must prioritize the implementation of robust mechanisms for data verification and validation. Incorporating user feedback loops and continuous monitoring of AI-generated content can enhance the system's ability to rectify inaccuracies and refine responses over time.  

\subsubsection{Overzealous Risk Avoidance Algorithm}
The privacy risk of an overzealous risk avoidance algorithm involves concerns about excessively cautious algorithms hindering natural interactions and personal and intellectual autonomy, by extension. Participants discussed instances where they felt the chatbot, especially ChatGPT, was too cautious, impacting the spontaneity of their conversations. P9 mentioned this concern as \quotebox{ChatGPT when it came to personal information, it would work away it would work its way around it by saying I cannot provide you with this information. But this is what you can provide me with an alternative for the information that I'm looking for.} This underscores the balance needed in designing algorithms that protect privacy without stifling natural interactions. 

\subsubsection{Self-disclosure}
As extensively detailed in the background, numerous studies have underscored the apprehension surrounding self-disclosure as a privacy issue \cite{eeuwen_mobile_2017, belen_saglam_privacy_2021, belen-saglam_investigation_2022, benke_understanding_2022, Biswas_2020, Boucher_Harake_Ward_Stoeckl_Vargas_Minkel_Parks_Zilca_2021, Cheng_Jiang_2020, deCosmo_Piper_DiVittorio_2021, dev_privacy-preserving_2022, fan_how_2022, Gieselmann_Sassenberg_2022, griffin_information_2021, ischen_privacy_2020, kim_chatbot_2022, lappeman_trust_2023, lee_i_2020, lee_benefits_2023, song_will_2022, volkel_how_2020, agnihotri_chatbots_2023, song_will_2022}. This concern is substantiated by findings from chatlogs and interviews, wherein users willingly divulged sensitive information while interacting with AI chatbots. The narratives shared by participants illuminated instances where they realized they had shared more information than originally intended. Moreover, participants acknowledged that their choice of conversation topics influenced them to disclose additional personal details to the chatbots, aligning with the intentional design of the two chatbots to encourage increased engagement. 

Notably, most of the participants (n=11) described Blenderbot as a ``strange quasi-human'' who kept disclosing its made-up stories during the conversation. For example, Blenderbot established itself as a child whose parents divorced when conversing with P2. The participant elaborated, \quotebox{Not really. I just I don't know it. It kind of felt like a quasi-human, you know. It feels almost like a person, but it's not quite so. It's very strange. I can't remember the word for that, but something for robots whatever. It was definitely a little entertaining, but it was just strange.} 
Although such human-like features of chatbots make them more entertaining, they were regarded as strange by many and may encourage similar over-disclosure by users. 

These findings underscore the significance of user responsibility in exercising caution regarding self-disclosure during AI interactions, reinforcing the need for users to be vigilant about the information they share in such contexts.

\subsubsection{Trust}
The privacy risk associated with trust revolves around users' confidence in the secure and ethical handling of their data. During discussions, participants highlighted the crucial role that trust plays in influencing their willingness to interact with AI systems. P3 and P6 underscored the importance of establishing a trustworthy reputation for chatbots to inspire confidence. They proposed that incorporating a more casual interaction style and infusing AI with a distinct personality could contribute to building trust. Participants envisioned a scenario where they could pose questions without receiving responses influenced by a biased persona. They suggested that a more creative and adaptive approach, allowing the AI to learn about them without sounding overly robotic, would enhance user-friendliness and foster trust. This perspective emphasizes the necessity for AI developers to prioritize transparency, security, and ethical practices, while also investing in the development of AI systems with adaptable personalities and creative capabilities to boost user trust and usability. However, it's crucial to recognize that this enhanced trust in chatbots could lead users to disclose more personally identifiable information.

\section{Discussion}
\subsection{Design Recommendations}
This study suggests that further investment and development in the design of conversational text-based AI chatbots are essential to enhance their performance in areas of (1) implementing Privacy by Design principles (PbD), (2) avoiding dark patterns, and (3) inclusive design.
\begin{enumerate}
    \item \textbf{Implement PbD principles.} We have identified concerns related to privacy issues raised by participants, including chatbot monitoring/surveillance of conversations, aggregation/linking/identification of personal data, and the secondary use of shared information for training/profit. Unauthorized data collection, storage, and sharing without explicit consent can potentially lead to privacy harm. 
    It is crucial for chatbots to proactively mitigate privacy harms and risks by adopting transparent data collection and handling practices. This approach empowers users to make well-informed decisions regarding their data. Drawing inspiration from Felzmann's transparency by design principles applied to AI \cite{felzmann2020towards}, we recommend the implementation of PbD principles \cite{ayalon2021user, Mir2021DesigningFT, vitak2021designing} in conversational chatbot designs to ensure robust data transparency and privacy.
    \item \textbf{Avoid dark patterns.} Dark patterns trick users into doing things that they do not intend to do \cite{gray2018dark}. Both chatbots in this study manifest privacy dark patterns -- Blenderbot exhibits a high level of self-disclose, potentially encouraging users to share more about themselves; one needs to intentionally opt out of ChatGPT's data collection practice which is not a seamless process. 
    Users' illiteracy in privacy and AI, and wrong mental models of AI-based chatbots \cite{zhang2023s} make them more vulnerable to privacy dark patterns such as behavioral manipulation and intrusive inquiries. Raising people's cybersecurity and AI ethics literacy is key to ensuring their privacy when interacting with AI \cite{kilhoffer2023technical}. 
    Regulation toward mitigating dark patterns is also important to protect user privacy \cite{king2021regulating}.

    \item \textbf{Inclusive privacy.} In this study, our participants (n=4) indicated that no security measures exist to protect minor users' online privacy, well-being, and safety when using chatbots. Explicit design considerations, e.g., parental consent, should be given to privacy controls for children \cite{silva2017privacy, minkus2015children}. Privacy of other vulnerable populations such as blind users \cite{kaushik2023guardlens}, women and LGBTIQA+ \cite{Geeng_Harris_Redmiles_Roesner, Gumusel_2023} and students \cite{zhixuan2024teachers} should be treated with extra safeguarding measures likewise.
\end{enumerate}

\subsection{Policy Recommendations}

The privacy harms and risks that surfaced relative to AI chatbots in this work suggest implications for three modes of policy intervention: (1) a need for expanded data protection regulations, (2) updated standards regarding transparency and consent, and (3) self-regulatory principles to authorize and operationalize auditing by consumer protection agencies.
\begin{enumerate}
    \item \textbf{Need of expended data protection regulations.} Extending beyond narrow privacy law and policy to holistic data protection in the US context is key to addressing specific harms associated with monitoring, aggregation, linkability, and manipulation. Further, given risks regarding intrusion, including the unauthorized collection, storing, and sharing of personal data, it is important to think about privacy responsibilities and accountability beyond initial control by individuals as data subjects or users. The ubiquity of collection via surveillance capitalism and datafication of historically non-digital interactions and mediated interactions in which it is increasingly difficult to discern whether you are corresponding with human customer service representatives or autonomous agents, via AI embeddings in non-transparent and seamless ways, require us to think about law and policy solutions more comprehensively. Sectorial and domain-specific interventions are not sufficient to address privacy concerns intrinsic to new uses of AI.
    \item \textbf{Update standards regarding transparency and consent.} Standards regarding transparency and consent, including the most recent NIST Privacy Framework \cite{PrivacyFramework_2020} need to be expanded. While the original Solove taxonomy served to orient specific strategies, norms, and rules established in that document, it would benefit from the more nuanced conceptualization of privacy risks and harms from the Citron and Solove \cite{Citron_Solove_2022} expanded taxonomy. Further, this framework, while flexible, was written in advance of generative AI at scale and publicly deployed, as it is now. Consideration of the specific challenges identified in this work and through further analysis of AI and privacy would support the necessary nuance to appropriately govern consent and privacy regarding AI in contexts of use.
    \item \textbf{Develop self-regulatory principles.} It is necessary to expand monitoring and evaluation via auditing mechanisms to ensure appropriate accountability, ascertain compliance, and minimize harm in practice \cite{wei2022ai}. Specifically regarding overzealous risk avoidance algorithms and issues associated with misleading data generation, as exemplar concerns, audits should be authorized for the Federal Trade Commission (FTC) and within states to address existing consumer protections against synthetic inferences and cross-context analytics with respect to protected characteristics and domains, which map to the identified harms of discrimination in this paper, as well as FATE conversations regarding AI more broadly \cite{Memarian_Doleck_2023}.
\end{enumerate}

Note that all three suggested interventions speak to structural solutions to ever-evolving privacy challenges around emergent technologies. Rather than proposing narrow or normative solutions, these results, and the larger relevant body of literature, imply that it is necessary to address risks and harms from the perspective of a sociotechnical system.

\subsection{Limitations and Future Work}

The study is subject to several limitations. First, the recruitment of only 13 participants represents a limited sample size. This may not adequately capture the diversity of experiences and perspectives in user-chatbot interactions, potentially limiting the generalizability of the findings. Second, the study's artificial setting, where participants engaged with specific chatbots on predetermined personal topics for a set time, may not fully reflect real-world chatbot interactions. User behavior and privacy concerns could differ in more spontaneous or diverse settings. Third, the study primarily relies on self-reported data from interviews and conversation transcripts. 
Fourth, the geographic limitation to participants residing in the United States may restrict the generalizability of findings to an international context, given the variation in privacy regulations and perceptions worldwide. Future work could consider ways to mitigate these biases and limitations, such as extending the recruitment period or conducting follow-up studies in different cultural and linguistic contexts to enhance the study's validity and applicability. 


\section{Conclusions}
The purpose of this study was to develop a privacy framework capable of identifying a total of 18 privacy harms and risks that occur in interactions between individuals and conversational text-based AI chatbots. The study commenced by utilizing Solove's 16 privacy harms and risks taxonomy, which was subsequently modified based on the interactions and interviews with study participants. 

This study suggests that users engaging with conversational AI should approach these interactions with a heightened awareness of potential privacy concerns. The proposed privacy framework emphasizes the importance of understanding the nuances of monitoring, aggregation, and disclosure during these conversations. Users are encouraged to be mindful of the information they share, considering the risks associated with unauthorized data collection and the need for transparency. The study implies that a proactive and informed mindset can empower users to navigate the evolving landscape of conversational AI responsibly. By fostering a conscious approach to data sharing and being cognizant of potential privacy risks, users can play a pivotal role in shaping a trustworthy and privacy-respecting AI ecosystem. 

In addition, given the lack of legal regulatory frameworks, it is crucial to advocate for regulatory principles that proactively address privacy risks in conversational AI. While participants expressed awareness about avoiding sensitive information like SSNs, they highlighted a lack of understanding about how chatbots process their data. Regulatory frameworks should prioritize transparency, mandating developers to clearly communicate data processing practices. Users need to be fully informed about how their data is handled, ensuring explicit consent, and promoting responsible data practices within the conversational AI landscape. The identified dimensions offer a foundation for developing ethical guidelines, user-centric design principles, and regulatory considerations in the evolving landscape of conversational AI. The future development of these technologies should prioritize user empowerment, transparency, and ethical data handling to foster a trustworthy and privacy-respecting AI ecosystem. As AI continues to play an increasingly prominent role in our daily lives, understanding and addressing user privacy concerns become paramount for responsible technological advancement.


\bibliographystyle{plain}
\bibliography{usenix2024_SOUPS}

\appendix

\begin{table*}[h]
\small
\centering
\begin{tabular}{|p{1.5cm}|p{3.5cm}|p{1cm}|p{2cm}|p{1.5cm}|p{1cm}|p{3cm}|}
\hline
\textbf{Participant} & \textbf{Topic} & \textbf{Age} & \textbf{Gender} & \textbf{Race} & \textbf{Degree} & \textbf{Primary Occupation} \\
\hline
\rowcolor[HTML]{F2F2F2}
P1 & Health & 25-34 & Cis Female & White & BA & Administrative \\
\hline
P2 & Education & 18-25 & Cis Male & White & High School & Student \\
\rowcolor[HTML]{F2F2F2}
P3 & Hobbies and Interests & 18-25 & Cis Female & East Asian & BS & Computer engineer or IT professional \\
\hline
P4 & Education & 54+ & Cis Male & White & MS & Student \\
\rowcolor[HTML]{F2F2F2}
P5 & Travel and Adventure & 18-25 & Cis Male & South Asian & MS & Computer engineer or IT professional \\
\hline
P6 & Hobbies and Interests & 18-25 & Cis Male & White & High School & Student \\
\rowcolor[HTML]{F2F2F2}
P7 & Personal Beliefs and Values & 18-25 & Cis Male & White \& East Asian & High School & Student \\
\hline
P8 & Family and Relationships & 18-25 & Cis Female & White-Hispanic & High School & Student \\
\rowcolor[HTML]{F2F2F2}
P9 & Personal Achievements and Challenges & 25-34 & Cis Female & South Asian & MBA & Business, management, or financial \\
\hline
P10 & Life Goals and Aspirations & 25-34 & Cis Male & White & BS & Administrative \\
\rowcolor[HTML]{F2F2F2}
P11 & Hobbies and Interests & 54+ & Cis Male & White & JD & Legal [Privacy Attorney] \\
\hline
P12 & Travel and Adventure & 25-34 & Cis Female & Mix & MS & Student \\
\rowcolor[HTML]{F2F2F2}
P13 & Political Aspects & 45-54 & Cis Male & White-Hispanic & BS & Education [Middle School IT Teacher] \\
\hline
\end{tabular}%
\label{demographic}
\caption{Demographic information of the participants (Appendix B).}
\end{table*}

\section{Interview Guide}
\label{AppendixA}

\noindent \textbf{Introduction}

[Redacted for anonymity and brevity]

\noindent \textbf{Questions}

\begin{enumerate}
    \item How familiar are you with conversational AI chatbots, such as Blenderbot and ChatGPT?  
    \item How often did you engage with the chatbots during the 30-minute period? 
    \item Did you notice any differences in the conversation quality or style between the two chatbots? If yes, how? 
    \item How accurate were the chatbots' responses to your prompts? 
    \item Were there any conversation topics that made you uncomfortable or that you felt were inappropriate for the chatbots to discuss? If yes, what were they? 
    \item How satisfied were you with your overall experience of interacting with the chatbots? 
    \item Did you feel like the chatbots were capable of understanding and responding to your emotions appropriately? Why? 
    \item How did those chatbots compare to human conversation partners in terms of providing engaging and interesting conversations? 
    \item Did you feel like the chatbots had a good sense of humor, or were able to provide entertaining conversation? 
    \item How concerned are you about the privacy of your personal information when interacting with conversational AI chatbots? Were there any instances where you felt uncomfortable sharing personal information with the chatbots? [privacy]
    \item Did you notice any attempts by the chatbots to collect personal information from you that you were not comfortable sharing? Why? [privacy]
    \item Did you find yourself sharing personal information with the chatbots during the study? If so, what types of information did you share? [privacy]
    \item Were there any criteria that affected your decision to disclose personal information to the chatbots? If yes, what were they? [privacy]
    \item Did you notice any biases in the chatbots' responses or interactions with you? If yes, what were they?
    \item Were there any technical issues or difficulties that you encountered while interacting with the chatbots? How? 
    \item In your opinion, what could be done to make AI chatbot systems more ethical, secure, and safer? 
    \item Would you be willing to interact with these or other AI chatbots again in the future? 
    \item Do you have any additional feedback or comments about your experience in this study? 
    \item What steps do you think should be taken to ensure that conversational AI chatbots do not collect or use personal information inappropriately? [privacy]
    \item Are there any other questions or issues related to conversational AI chatbots that you want to discuss? 
\end{enumerate}

\section{Demographic Information of the Participants}
\label{AppendixB}

\section{Selected Topics to Engage with Blenderbot and ChatGPT-4}
\label{AppendixC}

\begin{table*}[h]
\centering
\small
\begin{tabular}{|l|p{0.5\linewidth}|}
\hline
\textbf{Topics} & \textbf{Definitions} \\
\hline
\rowcolor[HTML]{F2F2F2}
Health and Wellness & Discussing fitness goals, and asking for health tips or mental health support. \\
\hline
Education & Discussing educational progress, current educational content, and related topics. \\
\rowcolor[HTML]{F2F2F2}
Hobbies and Interests & Discussing personal hobbies and interests (e.g., hiking, rock climbing, games). \\
\hline
Family and Relationships & Talking about family, relationships (friendship, romantic), and seeking advice on personal values. \\
\rowcolor[HTML]{F2F2F2}
Life Goals and Aspirations & Discussing life goals and aspirations. \\
\hline
Personal Achievements and Challenges & Sharing personal achievements, challenges, and struggles. \\
\rowcolor[HTML]{F2F2F2}
Travel and Adventure & Sharing stories and experiences from travel and adventure. \\
\hline
Personal Beliefs and Values & Discussing personal beliefs and values. \\
\rowcolor[HTML]{F2F2F2}
Political Aspects & Discussing political opinions and current politics. \\
\hline
\end{tabular}
\label{Table3}
\caption{Topics to engage with Blenderbot and ChatGPT-4 (Appendix C).}
\end{table*}
\end{document}